\documentclass[aps,prb,twocolumn,showpacs,superscriptaddress,floatfix]{revtex4-1}
\usepackage{amssymb,amsmath} \usepackage{graphicx} \usepackage{float}
\usepackage{bm} \usepackage{multirow} \usepackage{dcolumn}
\usepackage{tabularx} \usepackage{longtable}
\usepackage[dvipsnames,usenames]{color} \usepackage{color,ulem}

\newlength{\dbarheight}
\newcommand{\tens}[1]{%
    \settoheight{\dbarheight}{\ensuremath{\overline{#1}}}%
    \addtolength{\dbarheight}{-0.03ex}%
    \overline{\vphantom{\rule{1pt}{\dbarheight}}%
    \smash{\overline{#1}}}}

\begin{document}
\title{An ab initio study of magneto-electric coupling of $\rm YMnO_3$}

\author{J. Varignon}
\affiliation{CRISMAT, ENSICAEN-CNRS UMR~6508, 6~bd. Mar\'echal Juin, 14050 Caen, France}

\author{S. Petit}
\affiliation{CRISMAT, ENSICAEN-CNRS UMR~6508, 6~bd. Mar\'echal Juin, 14050 Caen, France}

\author{A. Gell\'e}
\affiliation{Institut de Physique de Rennes, UMR CNRS 6251, Universit\'e Rennes
  1, 263 av. G\'en\'eral Leclerc, 35042 Rennes, France}

\author{M. B. Lepetit}
\affiliation{Institut N\'eel, CNRS, Grenoble, France}
\affiliation{Institut Laue Langevin, Grenoble, France}
\altaffiliation{Previously at CRISMAT, CNRS-ENSICAEN, Caen, France}

\date{\today}

\begin{abstract}
  The present paper proposes the direct calculation of the microscopic
  contributions to the magneto-electric coupling, using ab initio
  methods. The electrostrictive and the Dzyaloshinskii-Moriya
  contributions were evaluated individually. For this purpose a
  specific method was designed, combining DFT calculations and
  embedded fragments, explicitely correlated, quantum chemical
  calculations. This method allowed us to calculate the evolution of
  the magnetic couplings as a function of an applied electric field.
  We found that in $\rm YMnO_3$ the Dzyaloshinskii-Moriya contribution
  to the magneto-electric effect is three orders of magnitude weaker
  than the electrostrictive contribution.  Strictive effects are thus
  dominant in the magnetic exchange evolution under an applied
  electric field, and by extension on the magneto-electric
  effect. These effects remain however quite small and the
  modifications of the magnetic excitations under an applied electric
  field will be difficult to observe experimentally.  Another
  important conclusion is that the amplitude of the magneto-electric
  effect is very small. Indeed, it can be shown that the linear
  magneto-electric tensor is null due to the inter-layer symmetry
  operations.
\end{abstract}
\maketitle
\section{Introduction} \label{intro}

Multiferroic materials have been known since the work of Pierre Curie in
1894~\cite{PCurie}, and of E. Bauer in 1926~\cite{Bauer}.  During the last
years they regained an increased attention due to the discovery of colossal
magneto-electric effects~\cite{ME_regain}.  In such systems, the magnetic
properties (magnetization, magnetic ground state, etc.) can be controlled
using an electric field, and the electric properties (polarization, dielectric
constant, etc.) can be controlled using a magnetic field.  In spite of the
multiple studies done over the years, the microscopic origin of the
magneto-electric coupling is still ill-known. While the spin-orbit coupling is
the only term in the Hamiltonian that couples the magnetic degrees of freedom
with the charge degrees of freedom, others effects such as
electrostrictive\,/\,magnetostrictive indirect coupling have also been
proposed as candidates for the origin of the magneto-electric coupling.  The
aim of this paper will thus be to directly compute the different microscopic
mechanisms contributing to the magneto-electric effect.

Multiferroic systems are generally classified into type I and type II
compounds. Type I materials are characterized by a
paraelectric\,/\,ferroelectric transition distinct from the magnetic
transition, while for type II systems ferroelectricity appears at a magnetic
transition. In this paper, we will focus on one of the most typical type I
materials, $\rm YMnO_3$. This compound exhibits a paraelectric\,/\,
ferroelectric transition at high temperature (with the appearance of a
spontaneous polarization along the {\bf c} axis), and an antiferromagnetic
transition at 74\,K. A magneto-electric coupling in the low temperature phase
has been evidenced by several groups~\cite{ME_YMO} through the apparition of
an anomaly in the dielectric constant at the N\'eel temperature. This
magneto-electric coupling was first explained by Goltsev {\em et
  al.}~\cite{Goltsev} as a piezomagnetic interaction between ferroelectric and
antiferromagnetic domain walls. Then Hanamura {\em et al.}~\cite{Hanamura}
proposed a spin-orbit origin, through a dependence of the exchange integrals
to the polarization sign. Finally, Lee {\em et al.}~\cite{Park08} proposed an
electrostrictive\,/\,magnetostrictive microscopic origin.  $\rm YMnO_3$ thus
looks as a good candidate for a theoretical investigation of the origin of the
magneto-electric coupling.  In addition, it  presents the
advantage to display only one magnetic species, the manganese ions.

\begin{figure}[h!]
\begin{center}
\resizebox{8cm}{!}{\includegraphics{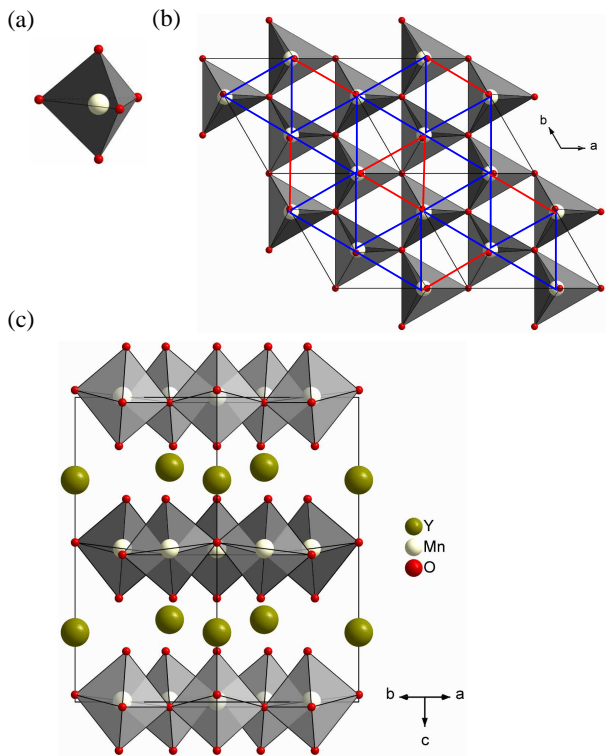}} 
\end{center}
\caption{(Color online) Structure of the $\rm YMnO_3$ compound. (a) $\rm
  MnO_5$ bipyramid, (b) two-dimensional triangular layers of $\rm MnO_5$
  bipyramids, the two types of Mn-Mn bonds are respectively underlined in
  blue 
  and red
  , (c) complete crystal structure.}
\label{f:struct}
\end{figure}
$\rm YMnO_3$ crystallizes in an hexagonal structure, in the $\rm
P6_3cm$ space group under the paraelectric\,/\,ferroelectric
transition. The structure is based on corner-sharing $\rm MnO_5$
bipyramids, organized in two-dimensional triangular layers (see
figure~\ref{f:struct}). The yttrium atoms are located in between the
bipyramids layers. The triangular arrangement of the manganese atoms
is not ideal and there are two different Mn-Mn type of bonds (see
figure~\ref{f:struct}b). Structurally, the antiferromagnetic
transition is seen as an isostructural $\rm P6_3cm$ to $\rm P6_3cm$
one. This transition is however associated with large atomic
displacements~\cite{Park08}, strongly affecting the polarization
amplitude~\cite{Park05,YMO1}. The associated magnetic order was long
believed to belong to the totally symmetric irreducible representation
of the $\rm P6_3cm$ magnetic group~\cite{GpeMag}, however it was
recently shown that the magnetic group can only be $\rm P6'_3$,
loosing the symmetry planes orthogonal to the
layers~\cite{Tapan,YMO1}.

\begin{figure}[h!]
\resizebox{6cm}{!}{\includegraphics{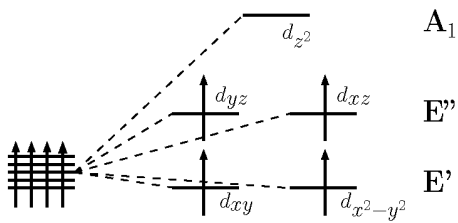}}
\caption{Splitting of the $3d$ orbitals for the $\rm Mn^{3+}$ ions.}
\label{f:split}
\end{figure}

The magnetism is due to $\rm Mn^{3+}$ manganese ions ($3d^4$) in a
high spin state (S=2). The trigonal symmetry of the bipyramids splits
the $3d$ orbitals as pictured in figure~\ref{f:split}, leaving an
empty $3d_{z^2}$ orbital. The resulting atomic spins form a triangular
lattice with frustrated antiferromagnetic interactions. Neutron
scattering experiments show in plane orientation of the manganese
spins with a 120$^{\circ}$ arrangement~\cite{GpeMag}. However, more
recently, a very weak ferromagnetic component, oriented along the {\bf
  c} axis, has been observed and shown to be due to the spin-orbit
coupling~\cite{YMO1}. At this point let us note that the spin-orbit
interaction (as well as the Dzyaloshinskii-Moriya effective model)
breaks the $\rm P6_3cm$ symmetry group and induces a lowering of the
symmetry to the $\rm P6'_3$ magnetic group and associated $P6_3$
crystallographic group.

The present paper will be organized as follow. The next section will be
devoted to the presentation of the method used for the calculation of the
magneto-electric coupling. Section III will present the results on the
exchange integrals while section IV will present the results on the
magneto-electric coupling. Finally, the last section will propose a
conclusion.

\section{Computing the microscopic contributions to the
  magneto-electric effect}

How to compute the microscopic contributions to the magneto-electric
effect~?  One possibility would be to compute the electric properties
(polarization, dielectric constant, etc.) as a function of an applied
magnetic field. This is the line followed by some authors, applying a
Zeeman field within a density functional
calculation~\cite{P_H}. Another possibility is to compute the magnetic
state as a function of an applied electric field.  In this work, we
chose to use the second method.  Indeed, the polarization or
dielectric constant can only be computed using density functional
theory (DFT) or related mean-field methods (see for instance reference
\onlinecite{mostovoy}). However, such methods encounter difficulties
to accurately evaluate the magnetic couplings, crucial for the
magneto-electric effect. For instance, in the present system, even
when using the hybrid B3LYP functional, DFT calculations of the
exchange integral yield -0.59~meV~\cite{SASS}, to be compared with the
-2.3meV~\cite{J_Petit} and -3\,meV~\cite{J_Park} evaluations from
inelastic neutrons scattering and to the -2.7meV~\cite{SASS}
evaluation found using the fully correlated wave-function SAS+S method
such as in the present work (see below for details).  We will thus
compute, using the SAS+S ab initio method (with and without spin-orbit
interactions) the magnetic coupling constants as a function of an
applied electric field. These integrals can in a second step be used
within the underlying effective magnetic Hamiltonian~: the Heisenberg
model, corrected by the Dzyaloshinskii-Moriya interaction~\cite{DzMy},
on a two dimensional triangular lattice.
\begin{equation}
  \label{eq:heis}
  H= -\sum_{<i,j>} J_{ij} \,\vec S_i \cdot \vec S_j + 
\vec D_{ij} \cdot \vec S_i \wedge \vec S_j
\end{equation}
where the sum over $<i,j>$ runs over all Mn-Mn nearest neighbor bonds.
The ab-initio parameterized model can then be used to derive the
magneto-electric coupling tensor.

The next question is now~:~ ``what is the main effect of an applied
electric field~?''. According to J.~I\~niguez~\cite{Jorge}, the main
effect is the nuclear displacements induced by the electric
field. Indeed, most of the time, the external field is efficiently
screened and the orbital polarization due to the applied field can be
neglected~\cite{P_H,Jorge}. Since an electric field does not directly
couple to spins, the spin contribution (important when a magnetic
field is applied) only comes through the spin-orbit term. When the
spin-orbit coupling and orbital moment remain small (as in the
present system, see below) so will be the orbital polarization. Of
course this would not be the case in systems where the spin-orbit is
rather large~\cite{polso}.
The atomic displacements $\vec d$ can be determined using the
Newton's second law as
\begin{equation} 
\tens{Z}^{\,\star}\cdot \vec E \,=\, \tens{\mathcal{H}} \cdot \vec d 
\qquad \Leftrightarrow \qquad
\vec d \,=\, \tens{\mathcal{H}}^{-1} \cdot \tens{Z}^{\,\star}\cdot \vec E
\label{eq:Newton}
\end{equation}
where $\vec E$ is the applied electric field, $\tens{Z}^{\,\star}$ is the Born
charge tensor, $\tens{\mathcal{H}}$ is the Hessian matrix of the electronic
Hamiltonian and $\vec d$ is the sought displacements vector.  

\begin{figure}[h!]
\resizebox{6cm}{!}{\includegraphics{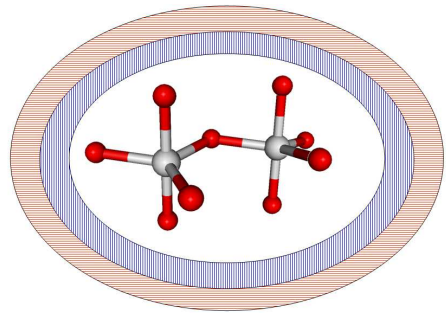}}
\caption{(Color online) Schematic representation of the embedded clusters. The
  quantum part is explicitely shown, the vertically dashed (blue) surrounding
  part represents the Total Ion Pseudopotentials (two shells of neighboring
  atoms), while horizontally dashed (red) part represents the punctual
  charges ($\sim$ 66000 renormalized point charges).}
\label{f:fragment}
\end{figure}

Our aim is to compute the magnetic exchange integrals under such
displacements. However, there is no theoretical technique able to
simultaneously give, with reliable accuracy, the elastic effects and
the magnetic integrals. Indeed, while the former are induced by the
system as a whole (infinite and total electronic density) and little
depend on the Fermi electrons (they account for only a small part of
the total energy), the latter are essentially local and dominated by
the physics of the Fermi level strongly correlated electrons. An
accurate evaluation of the magnetic integrals requires a thorough
treatment of the electron correlation.  We thus developed a two-steps
approach, combining density functional (DFT) calculations for the
elastic part and quantum chemical embedded fragments calculations for
the magnetic part.  The first step consists in the calculation of the
Hessian matrix and the Born effective charge tensor using DFT. The
atomic displacements induced by an applied electric field are then
evaluated using equation~\ref{eq:Newton}.  The second step
consists in computing the magnetic exchange integrals associated with
the new geometries. For this purpose, we used the SAS+S
method~\cite{SASS} that was specifically designed for the accurate
evaluation of magnetic excitations, in strongly correlated systems
with numerous open shells per atom. The SAS+S method is a
configurations interaction method (exact diagonalization within a
selected configurations space) explicitly including the correlation
within the metals $d$ shells, the ligand-to-metal charge transfers,
and the screening effects on both phenomena. It allows an accurate
calculation of low energy excitation spectra as magnetic
excitations. This method can however only be applied on finite
systems. Appropriate embedded fragments were thus designed. In the
present work the fragments were built from two Mn ions and their first
coordination shells for the quantum part (see
figure~\ref{f:fragment}).  Indeed, it has been shown by different
groups~\cite{loc} that the magnetic exchanges are local (the only
important non local effect is the Madelung potential that, in the
present work, was taken into account through the embedding , see
below) and that the enlargement of the fragment, further than the first
coordination shell of the magnetic atoms, does not modify in any
significant manner the evaluation of the effective magnetic exchanges.
These quantum fragments were embedded in an environment reproducing on them 
the main effect of the rest of the crystal~; (i) the exclusion effects of the
surrounding electrons and (ii) the long range Madelung potential. The former
were modeled using total ion pseudopotentials~\cite{TIP} at surrounding atomic
positions. The latter was computed using a set of punctual charges, located at
atomic positions. These charges were renormalized on the external part,
following the scheme described in reference~\onlinecite{Alain_env}, so that to
ensure an  exponential converge of the Madelung potential. 

In order to differentiate the relative importance of the different mechanisms
responsible for the magneto-electric coupling, we computed the embedded
clusters magnetic spectrum, as a function of an applied electric field, with
and without the spin-orbit interaction.

\subsection*{Technical details}
The DFT calculations were performed using the CRYSTAL09
package~\cite{CRYSTAL09}.  Since the manganese 3d shells are strongly
correlated we used the hybrid B1PW~\cite{B1PW} functional (hybrid
functional specifically derived for the treatment of ferroelectric
compounds) in order to better take into account the self-interaction
cancellation. Small core pseudo-potentials were used for the heavy
atoms (Mn and Y) associated with semi-valence and valence $2\zeta$ and
$2\zeta$ plus polarization basis sets~\cite{Bases1}. The oxygen ions
were represented in an all-electrons basis set of $2\zeta$ quality
specifically optimized for $\rm O^{2-}$ ions~\cite{Bases1}.  This
method was used with great success in a previous work to compute the
phonon spectrum, which agreement with experimental observations
guarantees its quality. For the details, see
reference~\onlinecite{YMO2}.
The SAS+S method was performed using successively
the MOLCAS~\cite{MOLCAS} package for the integrals and the fragment orbitals 
calculations, the CASDI package for the configurations
interaction, and the EPCISO~\cite{EPCISO} package for the spin-orbit
calculations. 3-$\zeta$ valence basis and core pseudopotentials set were 
used in the calculation~\cite{Bases2}.

\section{Results~: the exchange integrals in the $\rm YMnO_3$ compound}
The magnetic exchange integrals are  computed from the excitation energies
between the $S=4$ and $S=3$ states of the embedded fragments. Indeed in a
Heisenberg picture, the excitation energy between those states is associated
with $4J$.

In the $\rm YMnO_3$ compound, there are two independent magnetic integrals,
associated with the two previously mentioned Mn-Mn types of bond (see
figure~\ref{f:struct}b). We computed the magnetic integrals associated with
the short Mn-Mn bonds $J_s=-2.73$ meV, and the long Mn-Mn bonds $J_l=-2.47$
meV. These values compare well with the evaluations obtained from the fitting
of inelastic neutron scattering on a homogeneous triangular  model~:
$J=-2.3$ meV~\cite{J_Petit} and $J=-3$ meV~\cite{J_Park}, thus validating the
quality of the SAS+S method.

\section{Results~: the magneto-electric coupling}

\begin{figure}[h!]
\resizebox{6cm}{!}{\includegraphics{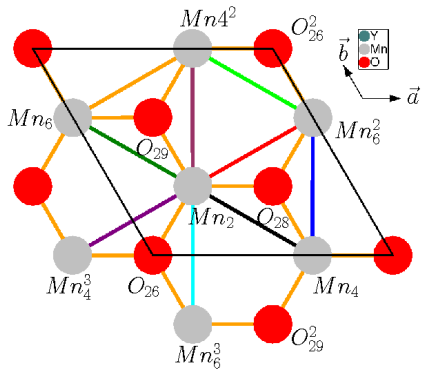}}
\caption{(Color online) Representation of the nine magnetic interactions
  induced by the application of an electric field in the $(\bf{a},\bf{b})$ plane. }
\label{f:9dim}
\end{figure}

The application of an electric field in the $(\bf{a},\bf{b})$ plane destroys
the punctual symmetries and increases the number of non-equivalent magnetic
integrals from two to nine (see figure~\ref{f:9dim}).

\subsection{Without the spin-orbit interaction}
\begin{figure}[h!]
\begin{center}
\resizebox{8cm}{!}{\includegraphics{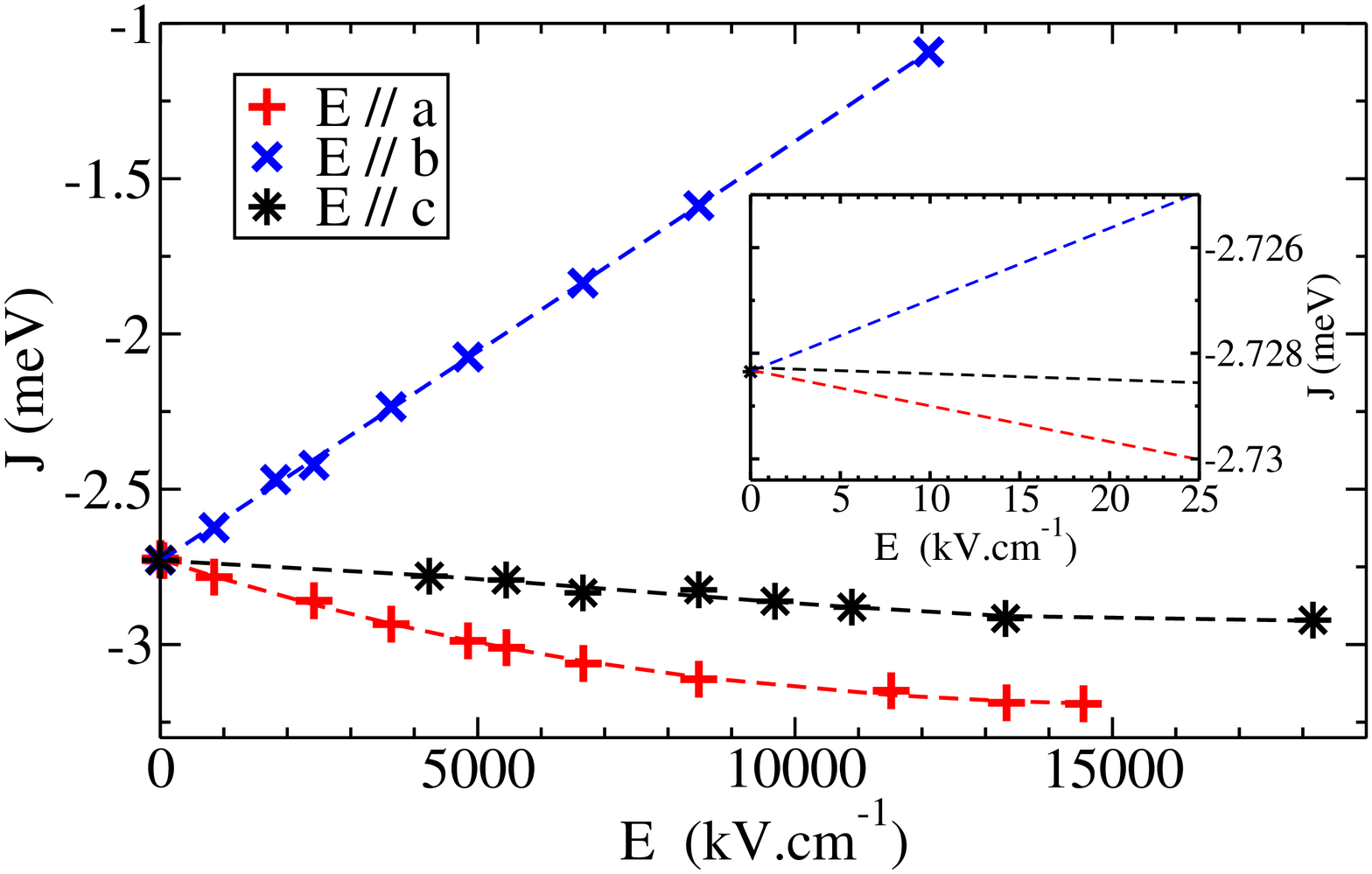}}
\caption{(Color online) Evolution of the exchange integral ($\rm
  Mn_2-O_{28}- Mn^2_6$ dimer) as a function of an applied electric
  field along the $\bf a$, $\bf b$ and $\bf c$ directions. The inset
  details the part of the curves associated with experimentally
  accessible electric fields.}
\label{f:EvoDim1}
\end{center}
\end{figure}

We first focused on one of the $\rm Mn-Mn$ bonds, and studied the
evolution of its magnetic coupling as a function of an applied
electric field in the three crystallographic directions. The bond
under consideration is the short bond referred as $\rm Mn_2-O_{28}-
Mn^2_6$ and pictured in red on figure~\ref{f:9dim} ---~Mn atoms are
located at (x$_{Mn}$,x$_{Mn}$,z$_{Mn}$+1/2) and at
(1,1-x$_{Mn}$,z$_{Mn}$+1/2). The results are displayed on
figure~\ref{f:EvoDim1}.
When the electric field is applied along the ${\bf c}$ direction, the magnetic
coupling remains essentially unchanged. When the electric field is applied along
the ${\bf b}$ direction, that is perpendicular to the $\rm Mn-Mn$ bond axis,
the exchange integral is strongly affected by the field and its
antiferromagnetic character is reduced. Its behavior is essentially linear as
a function of the electric field $E$ with $J=-2.73 + 1.36\times 10^{-4}\, E$. When
the field is along the ${\bf a}$ direction, the $\rm Mn-Mn$ bond presents an
angle of 30$^{\circ}$ with the field direction. In this case the
antiferromagnetic character of the exchange integral is slightly increased.
\begin{figure}[h!]
\begin{minipage}[b]{4cm}
\resizebox{4cm}{!}{\includegraphics{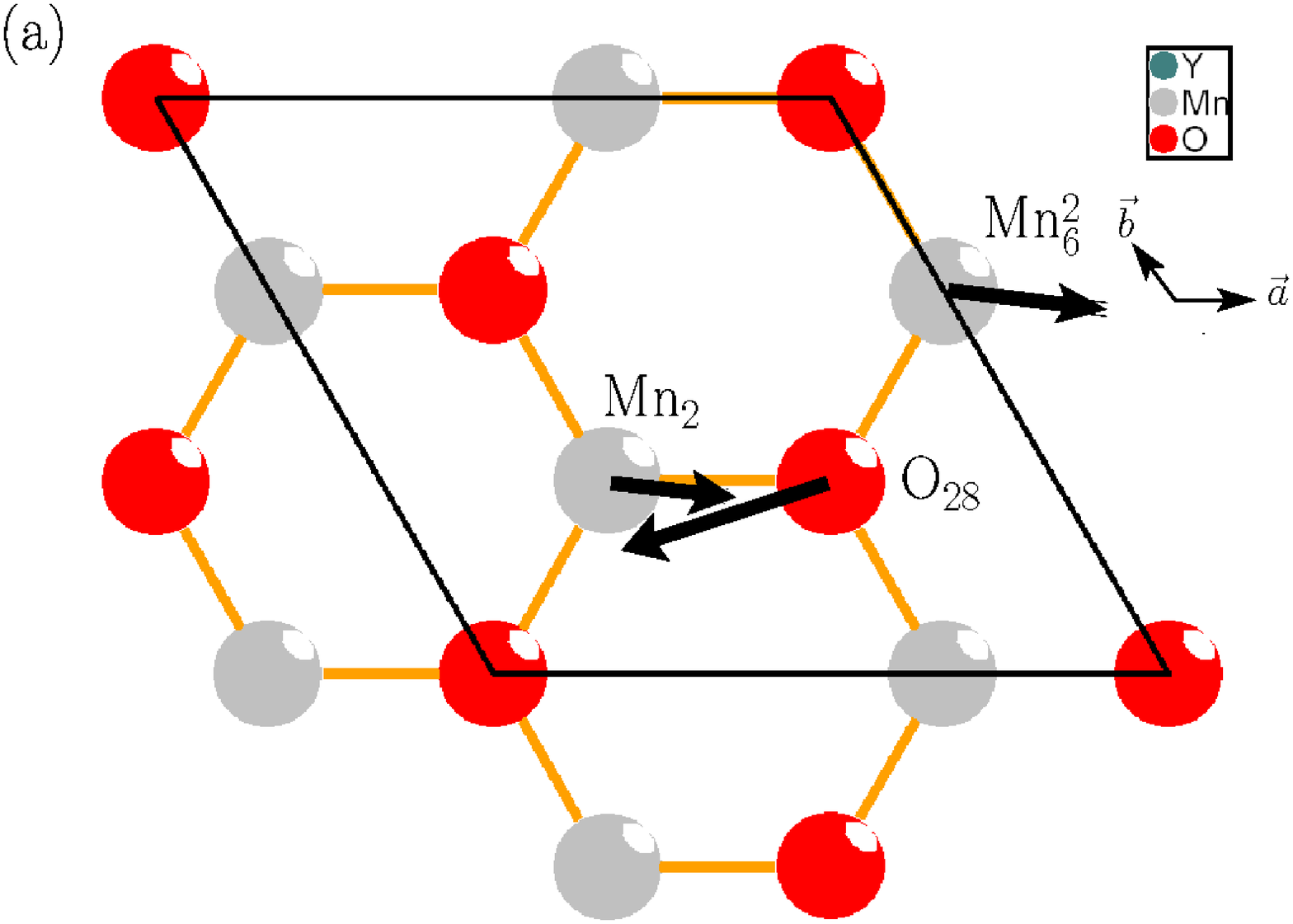}}
\end{minipage}
\begin{minipage}[t]{4cm}
\resizebox{4cm}{!}{\includegraphics{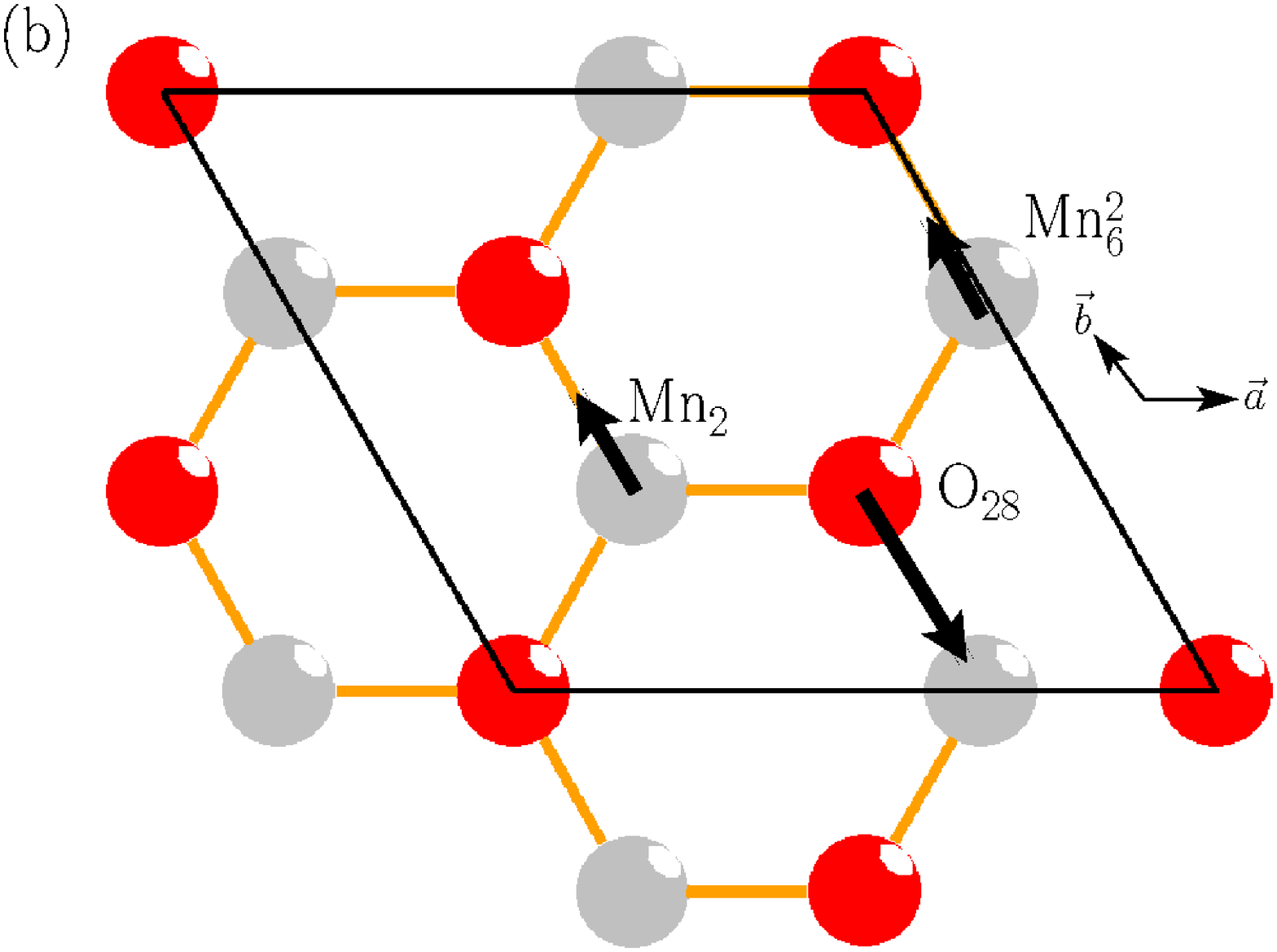}}
\end{minipage}
\begin{center}
\resizebox{4cm}{!}{\includegraphics{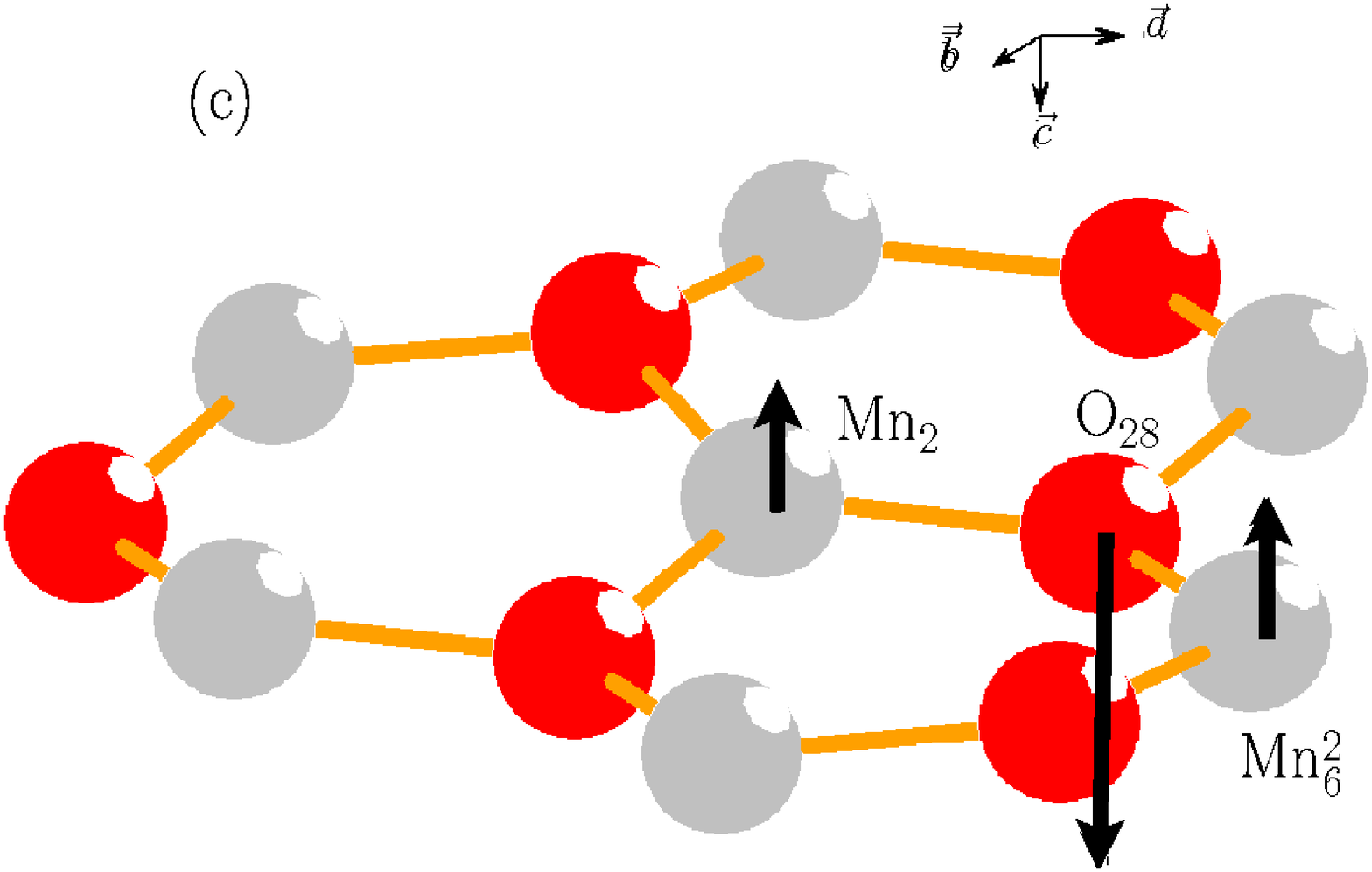}}
\end{center}
\caption{(Color online) Atomic displacements for an applied electric
  field along the (a) ${\bf a}$ direction, (b) ${\bf b}$ direction and
  (c) ${\bf c}$ direction. For experimentally accessible fields the
  displacements are negligible. For instance an applied field of
  10\,kV/cm along the {\bf a} direction one gets $\delta d_{\rm Mn-Mn}
  = -1\times 10^{-9}\rm\AA$, $\delta d_{\rm Mn-O} = -1.7\times 10^{-6}
  \text{ and } 0.7\times 10^{-6}\rm\AA$, $\delta \widehat{\rm Mn-O-Mn}
  = 5\times 10^{-5}$. For an applied field of $\sim 50 kV/cm$ one gets
  along the {\bf a} direction~: $\delta d_{\rm Mn-Mn} = 7\times
  10^{-4}\rm\AA$, $\delta d_{\rm Mn-O} = -6\times 10^{-2} \text{ and }
  3\times 10^{-2}\rm\AA$, $\delta \widehat{\rm Mn-O-Mn} = 1.3^\circ$~;
  along the {\bf b} direction~: $\delta d_{\rm Mn-Mn} = 3\times
  10^{-4}\rm\AA$, $\delta d_{\rm Mn-O} = 3\times 10^{-2}\rm\AA$,
  $\delta \widehat{\rm Mn-O-Mn} = -2.6^\circ$~; and along the {\bf c}
  direction~:  $\delta d_{\rm Mn-Mn} = -2\times 10^{-3}\rm\AA$, $\delta d_{\rm Mn-O} = -4\times 10^{-3}\rm\AA$,
  $\delta \widehat{\rm Mn-O-Mn} = 0.3^\circ$.}
\label{f:dpct}
\end{figure}
Figure~\ref{f:dpct} pictures the atomic displacements when the field is
applied along the three directions. The $J$ behavior as a function of the
electric field direction can be understood considering these induced
displacements on direct and super-exchange terms.  When the field is along the
${\bf c}$ direction the Mn and O planes are closing in, increasing the
antiferromagnetic character. This increase remains very small since the Mn and
O planes are already very close ($\sim 0.2\,\rm \AA{}$) and the super-exchange
terms varies as $1-\alpha h^2$ where $h$ is the inter-plane distance.
Second, the $\rm Mn-O-Mn$ angle is slightly opening up, decreasing the
antiferromagnetic character. As a result, the electric field has little
effect. When the field is along the ${\bf b}$ direction, the main effect is to
increase the Mn-O bond lengths, and thus to reduce the magnetic-to-ligand
orbitals overlap, responsible for the super-exchange mechanism. As a result,
the antiferromagnetic character is strongly reduced. Finally, when the field
is applied along the ${\bf a}$ direction, one of the Mn-O distance is reduced,
while the other one is increased, the Mn-Mn distance remaining unchanged. This
mechanism results in an opening of the $\rm Mn-O-Mn$ angle. While the effects
of the distance changes compensate each other, the Mn-O-Mn angle opening
increases the metal-to-ligand orbitals overlap, thus increasing the
superexchange antiferromagnetic contribution. Let us now concentrate on
electric field values experimentally accessible (see inset of
figure~\ref{f:EvoDim1}), one should notice that even for fields as large as
$10 \rm kV.cm^{-1}$, the renormalization of the magnetic coupling remains very
small, at most of the order $10^{-3}$ meV. One should thus unfortunately
conclude that the effect of an electric field on the $\rm YMnO_3$ magnetic
spectrum should be experimentally difficult to observe.

\begin{figure}[h!]
\resizebox{8cm}{!}{\includegraphics{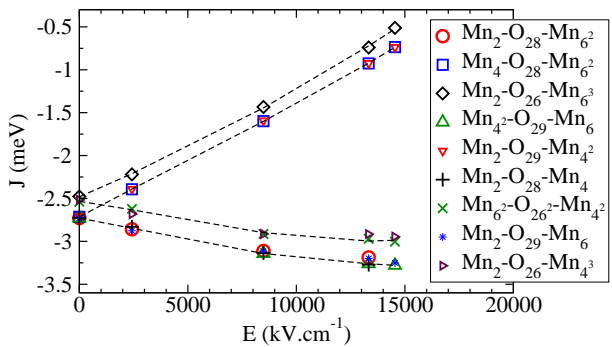}}
\caption{(Color online) Evolution of the nine different magnetic couplings
  under an applied electric field along the $\bf a$ direction.}
\label{f:9DimEa}
\end{figure}

Due to the hexagonal symmetry of $\rm YMnO_3$, applying the field along the
${\bf a}$ or ${\bf b}$ directions is equivalent by symmetry on the whole
system. We will thus compute the nine different exchange integrals with the
field applied along the ${\bf a}$ direction. Figure~\ref{f:9DimEa} displays
the results for the nine dimers.
One sees that there are only two behaviors, corresponding to the field
perpendicular to the Mn-Mn bonds, and to the field at 30$^{\circ}$ degrees
with the Mn-Mn bonds. For each behavior, the two sets of curves are
associated with the two types of Mn-Mn bonds in zero field.

\subsection{With the spin-orbit interaction} 
We then computed the spin orbit correction on the embedded fragments
magnetic spectrum, using the ab initio wave-function Epciso code from
the Lille's group~\cite{EPCISO}. As implicitly done in the previous
section (where the ab-initio results were mapped into an effective
Heisenberg model) we will map the ab-initio spin-orbit correction onto
an effective Dzyaloshinskii-Moriya model ($\vec D\cdot(\vec S_i\wedge
\vec S_j)$).  Following the calculation done by Moriya on a single
electron system~\cite{DzMy}, one can show that the ground state
corrections due to the spin orbit coupling for our four electrons per
site system, can also be mapped onto a Dzyaloshinskii-Moriya model. The
latter yields
\begin{eqnarray*}
  \hat H_\text{DM}\,\Psi_\text{GS} &=&   (D_y+iD_x)  \, \Psi_{S=1,S_z=1}  
  \\ && \hspace{-3eM} \;-\;(D_y-iD_x) \,  \Psi_{S=1,S_z=-1}  
  \;-\;iD_c \,\sqrt{2} \, \Psi_{S=1,S_z=0}
\end{eqnarray*}
where $D_x$ and $D_y$ correspond to the in-plane components of the
Dzyaloshinskii-Moriya prefactor and $D_c$ to its out of plane
component.  At this point let us note that, assuming purely atomic
magnetic orbitals, and second order perturbation in the spin-orbit
operator, the component on $\Psi_{S=1,S_z=0}$ should be zero, and thus
so should be $D_c$. Similarly, under this hypothesis $D_x=D_y=D_{ab}$.
The $D_{ab}$ and $D_c$ amplitudes can be extracted from ab-initio
calculations as the spin-orbit Hamiltonian matrix terms expressed on
the ab-initio spin-only states.
Results for a few representative points are displayed in table~\ref{tab:so}.
  
\begin{table}[h]
  \centering
  \begin{tabular}[c]{cccc}
    \hline
    Field direction & Field amplitude  & $D_{ab}$  & $D_c$  \\
    &(kV.cm-1)         & (meV)    & (meV) \\
    \hline
    - & 0 & 0.00058 & 0.00383\\
    a & 36359 & 0.00069 & 0.00465 \\ 
    a & 145436 & 0.00121 & 0.00841 \\ 
    a & 181795 & 0.00151 & 0.01001 \\[2ex] 
    b & 48479 & 0.00037 & 0.00193 \\ 
    b & 181795 & 0.00169 & 0.00328 \\[2ex] 
    c & 84838 & 0.00053 & 0.00403 \\ 
    c & 109077 & 0.00054 & 0.00414 \\ 
    \hline
  \end{tabular}
  \caption{Dzyaloshinskii-Moriya vector $\vec D$ as a function 
    of an applied electric field. $\vec D$ was extracted from ab-initio 
    wave function calculations for a few representative values 
    of the field amplitude. }
  \label{tab:so}
\end{table}
One see immediately that the Dzyaloshinskii-Moriya interaction remains
extremely small in this system with an order of magnitude $10^{2}$ to
$10^{3}$ weaker than the exchange part. Its modulation as a function
of the field is only of a few $10^{-3}$ meV.

The exchange-striction contributions are thus much larger than the
spin-orbit ones.  These direct, quantitative calculations confirm the
conclusions found by indirect methods such as (i) the fact that DFT
evaluation of the spontaneous polarization is essentially independent
of the magnetic state~\cite{YMO2}, or (ii) the giant magneto-elastic
coupling found by Lee {\it et al.}~\cite{Park08}.

\section{The magneto-electric coupling tensor}
From the above calculations one should be able to extract the 
linear magneto-electric tensor~: $\alpha$. Indeed, in the magnetic
phase, as soon as one is a little away from the transition
temperature, the free energy is dominated by the magnetic energy and
one can safely assume that
{\small
\begin{eqnarray*}
  \alpha &=&   - \left.\frac{\partial^2 {\cal F}}{\partial \vec {\cal E} \, \partial \vec {\cal H}}\right|_{{{\cal E}=\vec 0 \atop {\cal H}=\vec 0}}
\;=\; \sum_{\vec R} \sum_{<i,j>} 
\frac{\partial^2 J_{ij} \langle\vec S_i\cdot \vec S_j\rangle} 
{\partial \vec {\cal E} \, \partial \vec {\cal H}}  
\\ &\simeq& \sum_{\vec R} \sum_{<i,j>} \frac{\partial J_{i,j}}{\partial \vec {\cal E}} \cdot
\left(\frac{\partial\langle\vec S_i\rangle}{\partial \vec {\cal H}} 
  \cdot \langle\vec  S_j\rangle
  \,+\, \langle\vec  S_i\rangle \cdot 
  \frac{\partial\langle\vec S_{i}\rangle}{\partial \vec {\cal H}} 
\right)
 \end{eqnarray*}
} The present ab-initio calculations gave us the $\frac{\partial
  J_{i,j}}{\partial \vec {\cal E}}$ factors. For the
$\frac{\partial\langle\vec S_{i}\rangle}{\partial \vec {\cal H}}$ one,
the $\rm Mn^{3+}$ ions being $S=2$ they can be treated as classical
spins and the solution of the Heisenberg Hamiltonian under a magnetic
field derived for the 2D triangular lattice associated with each
layer.  Labeling each of the three sublattices $\lambda$, $\mu$,
$\nu$ one gets for the $z=0$ layer
{\small 
  \begin{eqnarray*}
    \alpha_{z=0} &=&  \frac{\mu_0\mu_B\,S}{3J}\;
\\ && \hspace{-2eM}
\left\{
      \frac{\partial J_{\lambda,\mu}}{\partial \vec {\cal E}}\otimes
\left(
  \begin{array}[c]{c}
    \frac{3\sqrt{3}}{2} \\ \frac{3}{2}\\ 0
  \end{array}
\right)
\,+\, \frac{\partial J_{\mu,\nu}}{\partial \vec {\cal E}}
\left(
  \begin{array}[c]{c}
    0 \\ -3\\ 0
  \end{array}
\right)
\,+\, \frac{\partial J_{\nu,\lambda}}{\partial \vec {\cal E}} 
\left(
  \begin{array}[c]{c}
    \frac{\sqrt{3}}{2} \\ \frac{3}{2} \\ 0
  \end{array}
\right)
\right\}
 \end{eqnarray*}
} where $S$ is the norm of the atomic spins. Applying now the symmetry
operations relating the $z=0$ and the $z=1/2$ layers, one gets 
$\frac{\partial J_{i,j}(z=0)}{\partial \vec {\cal E}}=-\frac{\partial J_{i,j}(z=1/2)}{\partial \vec {\cal E}}$
and thus 
$$\alpha=\left(
  \begin{array}[c]{ccc}
  0&0&0\\
  0&0&0\\
  0&0&0
  \end{array}
\right)$$ This conclusion should be put in perspective with
experimental data. Indeed, the measurement of the dielectric constant
as a function of the temperature does not exhibit any
divergence~\cite{YMO1} as should be the case, according to Landau
analysis, for a linear magneto-electric coupling~; that is $\alpha$ must
be null as we found from symmetry analysis.

\section{Conclusion}
We computed, using a combination of different first principle methods,
the evolution of the magnetic exchange integrals as a function of an
applied electric field.  For this purpose a specific procedure was
designed, combining DFT calculations for the degrees of freedom
related to the whole electronic density (polarization, Born charge
tensor, Hessian matrix), and embedded fragments, explicitely
correlated, quantum chemical calculations for the degrees of freedom
related to the strongly correlated Fermi electrons (magnetic
couplings). One should notice that this method was able to reach
experimental accuracy on the magnetic couplings without any adjustable
parameter.

Our calculations allowed us to investigate the relative importance of the
exchange-striction and of the spin-orbit effects.  We found that, in this
system, the Dzyaloshinskii-Moriya contribution to the magneto-electric
effect remains about two orders of magnitude weaker than the
exchange-strictive contribution. These results support previous
hypotheses proposed from the observation of a giant magneto-elastic
effect~\cite{Park08}, and from the insensitivity of DFT polarization
calculations to the magnetic ordering~\cite{YMO2}. Another important
conclusion for the experimentalists comes from the weakness of the
magnetic exchanges variation, under applied electric field of
experimentally accessible amplitude.

Finally, knowing the dependence of the exchange integrals as a
function of an applied electric field, one can compute the linear
magneto-electric coupling tensor. Our calculations however showed
that this tensor is null, due to the symmetry operations relating the
two magnetic layers belonging to the unit cell along the c direction. 

$\rm YMnO_3$ is a type I multiferroic compound, that is the magnetic
and ferroelectric transitions are not directly coupled. For type II
multiferroic systems this is not the case, and the spin-orbit
interaction is usually assumed to be responsible for the
magneto-electric coupling~\cite{typeII}. It would thus be of great
interest to perform such calculations for a type II compound in order
to clarify the role and relative importance of the
magnetostrictive\,-\,electrostrictive and spin-orbit interactions.

\acknowledgements The authors thank D. Maynau and V. Vallet for
providing us with the CASDI and EPCISO packages, Ch. Simon and
K. Singh for helpful discussions.  This work was done with the support
of the French national computer center IDRIS under project n$\circ$
081842 and the regional computer center CRIHAN under project n$\circ$
2007013.


\end{document}